\documentclass[preprint,trackchanges]{aastex62}

\newcommand{\speed}[1]{#1 km~s${}^{-1}$}

\usepackage{lineno}
\usepackage{url}
\usepackage{hyperref}
\shorttitle{Formation of Fan-spine Magnetic Topology}
\shortauthors{Duan et al.}

\begin{document}
\title{Formation of Fan-spine Magnetic Topology through Flux Emergence and Subsequent Jet Production}

\correspondingauthor{Hui Tian}
\email{huitian@pku.edu.cn}

\author[0000-0001-9491-699X]{Yadan Duan}
\affil{School of Earth and Space Sciences, Peking University, Beijing, 100871, People's Republic of China}
\author[0000-0002-1369-1758]{Hui Tian}
\affil{School of Earth and Space Sciences, Peking University, Beijing, 100871, People's Republic of China}
\author[0000-0001-7866-4358]{Hechao Chen}
\affil{School of Physics and Astronomy, Yunnan University, Kunming 650500, People’s Republic of China}
\author[0000-0001-9493-4418]{Yuandeng Shen}
\affil{Yunnan Observatories, Chinese Academy of Sciences, Kunming, 650216, China}
\author[0000-0001-5657-7587]{Zheng Sun}
\affil{School of Earth and Space Sciences, Peking University, Beijing, 100871, People's Republic of China}
\author[0000-0003-4804-5673]{Zhenyong Hou}
\affil{School of Earth and Space Sciences, Peking University, Beijing, 100871, People's Republic of China}
\author[0000-0001-7693-4908]{Chuan Li}
\affil{School of Astronomy and Space Science, Nanjing University, Nanjing 210023,China}
\affil{Key Laboratory for Modern Astronomy and Astrophysics (Nanjing University), Ministry of Education, Nanjing 210023, China}

\begin{abstract}

Fan-spine magnetic structure, as a fundamental three-dimensional topology in magnetic reconnection theory, plays a crucial role in producing solar jets. However, how fan-spine configurations form in the solar atmosphere remains elusive. Using the Chinese H$\alpha$ Solar Explorer (CHASE) and the Solar Dynamics Observatory (SDO), we present a case study on the complete buildup of fan-spine topology driven by flux emergence and the subsequent jet production. Two fan-spine structures and the two associated null points are present. Variations in null-point heights and locations were tracked over time during flux emergence. The north fan-spine structure is found to be created through magnetic reconnection between the newly emerged flux and the background field. Gentle reconnection persistently occurs after formation of the north fan-spine structure, resulting in weak plasma outflows. Subsequently, as flux emergence and magnetic helicity injection continue, the formation and eruption of mini-filaments after reconnection at the quasi-separatrix layer between the two nulls trigger three homologous jets. The CHASE observations reveal that the circular flare ribbon, inner bright patch, and remote brightening all exhibit redshifted signatures during these jet ejections. This work unveils the key role of flux emergence in the formation of fan-spine topology, and highlights the importance of mini-filaments for subsequent jet production.
\end{abstract}


\section{Introduction} \label{sec:intro}
\par
Solar jets are ubiquitous, collimated and transient plasma ejections that launch from the lower solar atmosphere along the magnetic field. They are indicators of energy release events in the solar atmosphere and play a crucial role in the mass and energy transport into the upper atmosphere \citep[e.g.,][]{2014Sci...346A.315T,2016SSRv..201....1R,2021RSPSA.47700217S}. Solar jets can be observed simultaneously but not strictly co-spatial at different wavelengths, such as  X-ray, extreme-ultraviolet (EUV) and H$\alpha$, reflecting their characteristic of exhibiting both hot and cool components \citep[e.g.,][]{2007A&A...469..331J,2010ApJ...720..757M,2012ApJ...745..164S,2017ApJ...851...67S}. Several studies have provided evidence supporting that the cool component of a jet is from the small eruptive filament confined within the base of jet \citep[e.g.,][]{2010ApJ...720..757M,2012ApJ...745..164S,2015Natur.523..437S,2018ApJ...863..180W,2019ApJ...881..132D,2020ApJ...902....8C,2023ApJ...953..148S,2023ApJ...953..171H}. The onset of a solar jet is often closely associated with magnetic flux emergence and cancellation \citep[e.g.,][]{2007A&A...469..331J,2015ApJ...815...71C,2016ApJ...833..150L,2017ApJ...844..131P,2018ApJ...869...78C,2019ApJ...887..220Y,2019ApJ...887..239Y,2021ApJ...923...45Z,2021ApJ...918L..20H,2023ApJ...942...86Y}.  Using spectroscopic data, previous observational works focused primarily on the plasma motion during jet eruptions, such as the simultaneous blueshifted and redshifted emissions on either side of the jet body, which can be explained by the rotational motion of solar jets \citep[e.g.,][]{2011A&A...532L...9C,2015ApJ...801...83C,2019ApJ...883...52R,2021MNRAS.505.5311K}. 
\par
Recently, more and more numerical simulations \citep[e.g.,][]{2009ApJ...704..485T,2009ApJ...700..559M,2009ApJ...691...61P,2010ApJ...714.1762P,2017Natur.544..452W,2018ApJ...852...98W} and observational studies \citep[e.g.,][]{2012ApJ...760..101W} have revealed that solar jets are closely related to the magnetic reconnection occurring in the three-dimensional (3D) fan-spine null point topology. The fan-spine structure consists of a dome-shaped fan separatrix surface below a null point (a point in space where the magnetic field vanishes), and the inner and outer spines passing through the null point \citep[e.g.,][]{2002A&ARv..10..313P}. The null point within the fan-spine magnetic topology system provides a highly favorable site for the occurrence of magnetic reconnection \citep[e.g.,][]{2011A&A...533A..78P}. In addition, many coronal mass ejections (CMEs) \citep{2021ApJ...907...41K}, white-light flares \citep{2018ApJ...867..159S,2018ApJ...854...64S}, small-scale coronal bright points \citep{2012ApJ...746...19Z}, ultraviolet bursts \citep{2017A&A...605A..49C,2019ScChE..62.1555C}, as well as macrospicules \citep{2023ApJ...942L..22D} are also suggested to occur in the fan-spine magnetic field configuration. Observationally, within the framework of fan-spine structure, the occurrence of a solar eruption often manifests as three flare ribbons. These include an inner bright patch encircled by a circular ribbon\citep[e.g.,][]{2015ApJ...812L..19L,2018ApJ...859..122L,2019ApJ...883...47L,2019ApJ...871..105Z,2022ApJ...926L..39D,2023A&A...674A.154M,2024arXiv240116101Z}, which signifies the footpoint positions of the inner spine and the dome-shaped fan structure. Additionally, a remote brightening indicates its association with the far-reaching end of the outer spine \citep[e.g.,][]{2009ApJ...700..559M,2013ApJ...778..139S,2015ApJ...804....4K,2015ApJ...806..171Y,2017ApJ...851...30X,2018ApJ...867..159S,2019ApJ...874..146H}. Recently, \citet{2017Natur.544..452W,2018ApJ...852...98W} conducted a high-resolution MHD simulation based on an embedded-dipolar model \citep{2009ApJ...691...61P,2010ApJ...714.1762P}, showing a breakout mechanism for solar eruptions in the fan-spine topology. However, this model did not explain the formation and evolution of the fan-spine configuration. 
\par
The only one numerical simulation work on the formation of a fan-spine structure was presented by \citet{2009ApJ...704..485T}. They used the flux emergence model to simulate the interaction between a magnetic flux rope and pre-existing fields, ultimately resulting in the formation of a 3D fan-spine configuration. However, very limited solar-limb observations can be regarded as supporting evidence of this process \citep[e.g.,][]{2009ApJ...704..485T,2011ApJ...728..103L,2016ApJ...819L...3Z}. Without reliable on-disk photospheric magnetic field measurements, the formation process of the fan-spine structure cannot be investigated in detail. Here, we perform an analysis of a textbook example of a fan-spine magnetic configuration that occurred in an ephemeral region on the solar disk. With joint observations from the Chinese H$\alpha$ Solar Explorer (CHASE; \citealt{2022SCPMA..6589602L}) and the Solar Dynamics Observatory (SDO; \citealt{2012SoPh..275....3P}), we provide, for the first time, the unambiguous evidence for the formation of 3D fan-spine magnetic field topology and subsequent jet production.

\section{DATA AND Methods}\label{sec:DATA}
The data of both SDO and CHASE we analyzed were taken on 2023 June 1 at N16E12, in a solar ephemeral region. The H$\alpha$ Imaging Spectrograph (HIS; \citealt{2022SCPMA..6589605L}) on board CHASE provides, for the first time, the seeing-free H$\alpha$ spectroscopic and imaging observations of the full solar disk in the wavelength range of 6559.7--6565.9 \AA~\citep{2022SCPMA..6589602L,2022SCPMA..6589603Q}. This enables us to study the detailed plasma motions during the eruption of the fan-spine jets. The temporal cadence, spatial, and spectral pixel sizes of CHASE imaging spectral data are 60 s, 0.52\arcsec , and 0.024 \AA~, respectively. The Atmospheric Imaging Assembly (AIA; \citealt{2012SoPh..275...17L}) and the Helioseimic and Magnetic Imager (HMI; \citealt{2012SoPh..275..207S}) on board SDO provide the extreme-ultraviolet (EUV) coronal response and photospheric magnetograms during the entire process of the fan-spine formation, respectively. The AIA(HMI) observations have a time cadence of 12 (45) s and a pixel resolution of 0\arcsec.6 (0\arcsec.5), respectively.
\par
The magnetic field plays a pivotal role in driving solar activity, yet the routine measurement of the coronal magnetic field continues to be a significant challenge \citep[e.g.,][]{2020ScChE..63.2357Y,2020Sci...369..694Y,2023RAA....23b2001C}. In this study, to characterize the 3D magnetic field of the fan-spine system, we chose the line-of-sight (LOS) HMI magnetogram as the bottom boundary for the local potential field extrapolation with a code provided by SolarSoftWare (SSW). We identified and tracked the null-points using the method introduced by \citet{2021ApJ...923..163P}. The calculation region \textbf({x=[-334\arcsec,-118\arcsec], y=[185\arcsec,399\arcsec]) is the same as that of the local potential field extrapolation, }a cubic box of 216$\times$214$\times$101 uniformly spaced grid points with $\bigtriangleup$x=$\bigtriangleup$y=$\bigtriangleup$z=0\arcsec.5. In this region, there are approximately 200 magnetic null points, and most of them are located at very low heights near the photosphere. Valid null points were defined as those above the middle chromosphere ($\sim$0.8 Mm). Within the region of our interest \textbf({x=[-211\arcsec,-171\arcsec], y=[254\arcsec,292\arcsec])}, only two magnetic null points meet this criterion and were named the north and south null points, respectively. The 3D coordinates of these two null points were obtained every 720 seconds.

\par
Additionally, the magnetic helicity injection in the ephemeral region can be calculated using this equation \citep{1984JFM...147..133B,2012ApJ...761..105L}:
\begin{math}
dH/dt=2\int_{S}^{}(\textbf{A}_{p}\cdot \textbf{B}_{t})V_{\bot n} dS - 2\int_{S}^{}(\textbf{A}_{p}\cdot \textbf{V}_{\bot t})B_{n} dS, 
\end{math}
where the vector potential of the potential field $B_{p}$ is represented by $A_{p}$, $B_{n}$( $B_{t}$) represents the tangential (normal) magnetic fields; $V_{\bot t}$ and $V_{\bot n}$ are the tangential and normal components of the velocity perpendicular to the filed lines ($V_{\bot}$), respectively. This velocity ($V_{\bot}$) can be computed according to 
\begin{math}
\textbf{V}_{\bot}=\textbf{V}-(\textbf{V}\cdot \textbf{B}/ B^{2})\textbf{B},
\end{math} where \textbf{V} is derived by the Differential Affine Velocity Estimator for Vector Magnetograms (DAVE4VM; \citealt{2008ApJ...683.1134S}) method. 

\par
The Doppler velocity of the H$\alpha$ line can be obtained by using the equation 
\begin{math}
v_{dop}=(\lambda_{1}-\lambda_{0})/\lambda_{0}\cdot{c},
\end{math}
where ${c}$ is the light speed, $\lambda_{0}$ is the reference wavelength. $\lambda_{1}$ is the line center derived by the weight-reverse-intensity method \citep[e.g.,][]{2016ApJ...816...30S}. The reference line center was derived by averaging the line centers over a selected quiet region near the ephemeral region.

\section{RESULTS} \label{sec:OBS}
Figure 1 presents an overview of the ephemeral region, accompanied by three homologous jets. During the eruption process of the jets, a remote brightening, a circular flare ribbon, as well as a core brightening surrounded by the circular flare ribbon can be observed, as shown in Figure 1 (A) and (E). These typical observational characteristics are consistent with previous observations of solar jet eruptions under the fan-spine framework \citep[e.g.,][]{2009ApJ...700..559M,2012ApJ...760..101W}.  Additionally, we investigated the evolution of the photospheric magnetic field within the region of interest. Figure 1 (C) and (D) show the changes in the HMI LOS magnetogram before the first jet eruption, revealing a significant increase in the negative-polarity magnetic flux. Such intrusion of negative flux into the pre-existing positive flux provides an ideal condition for the formation of a fan-spine topology \citep{2009ApJ...704..485T}. 
\par
\subsection{Evolution of null points}
The temporal variation of the negative magnetic flux within the white box marked in Figure 1 (A) is presented in Figure 2 (C).  Starting from 00:30 UT, there is a continuous flux emergence of negative polarity, with a dramatic increase around 04:34 UT.  The north null point was found to be present at 4.1 Mm above the photosphere at around 05:17 UT.  And a south null point is present at the height of 0.9 Mm at 04:58 UT. Before the first jet eruption ($\sim$ 06:38 UT), the north null point exhibits an upward trend and reaches its highest height of 5.5 Mm at 06:34 UT. Following the ejections of the three homologous jets, as indicated by the three vertical dashed lines in Figure 2 (C), the north null point descends to a lower height (1.56 Mm), compared to its initial height. The south null point maintains a height of $\sim$1 Mm before the jet eruption, from 04:58 UT to 06:34 UT. After 06:34 UT, the south null point also rises and disappears after reaching its maximum height of $\sim$2.5 Mm at 07:58 UT. Figure 2 (A) and (B) shows the trajectory of the magnetic null points with time, revealing their evolving spatial locations due to flux emergence. By employing a potential field extrapolation, we obtained the magnetic field configurations of the two fan-spine structures at three different times. These times correspond to the periods preceding the formation of the north fan-spine structure (Figure 2 (D) and (E)), when the north null point reaches its highest height (Figure 2 (F) and (G)), and subsequent to the jet eruptions (panels (H) and (I)), respectively. The blue lines represent the small emerging loops, while the pink lines represent the magnetic field lines formed after the interchange reconnection between the emerging loops and the surrounding open field lines. Subsequently, the blue and pink lines evolve into the inner spines of the north fan-spine magnetic field configuration. The formation of the south fan-spine likely also results from flux emergence, as the south null presents above 0.8 Mm also immediately after the dramatic increase of negative flux. However, from the available observations it is unclear whether it is caused by reconnection between the emerging and background fluxes. The south fan-spine structure can be seen from Figure 2 (F)$-$(G). Figure 2 (H) and (I) display the eventual morphology of the north fan-spine structure after three homologous jet ejections. Combining Figure 2 (A)$-$(B) and (H)$-$(I), it can be seen that the north null point has drifted to the northern edge of the emerging region. The disappearance of the south null point after 07:58 UT may be caused by the cancellation of negative flux in this region with the background flux, which results in destruction of the south fan-spine structure. Here we are most interested in the formation and dynamics in the north fan-spine structure. In the subsequent sections, unless otherwise specified, “fan-spine” refers to the north fan-spine structure, and “null point” refers to the north null point.
\par

\par
\subsection{Formation of the north Fan-Spine Structure}
Figure 3 clearly displays the reconnection signatures accompanying the formation process of the north fan-spine structure in the EUV 171 \AA~channel. At the beginning, beneath a coronal plume region, a set of small loops emerges, as shown in Figure 3 (A). As flux emergence continues, the newly emerged small loops gradually approach and interact with the coronal plume through an interchange reconnection process (see Figure 3(B)$-$(C)). As a result, a set of newly formed closed loops (marked by the pink arrow) and some open loops (denoted by the blue arrow) appear in Figure 3 (C). As such an interchange reconnection process proceeds, a spider- or anemone-like fan dome begins to form. By 05:17 UT, the moment of the first emergence of magnetic null point, a typical fan-spine configuration is created, with a relatively axisymmetric nature, as shown in Figure 3 (E). Following the null point emergence, plasma outflows can be observed to move bidirectionally along the outer spine of the fan-spine system, as denoted by the white arrows in Figure 3 (F)$-$(G). Consistent with the null-point drifting mentioned in Section 3.1, the fan-spine magnetic field system becomes notably non-axisymmetric due to the persistent photospheric flux emergence.  Meanwhile, a bright plasma sheet develops near the null-point site at $\sim$06:10 UT.

\par
Figure 4 illustrates the accumulation of magnetic helicity in the emerging region, and the occurrence of sustained magnetic reconnection at the null point of the fan-spine system. By integrating the helicity injection flux in the emerging region over time, we obtained the accumulated magnetic helicity through the bottom boundary within the volume of interest. The calculating time interval adequately encompasses the formation process of the fan-spine configuration. As displayed in Figure 4 (A), the calculated result shows a significant increase in the magnetic helicity accumulation from 04:34 UT. Such a sudden increase is clearly related to the substantial emergence of magnetic flux, as displayed in Figure 2 (C). It should be noted that the magnetic flux, magnetic helicity, and AIA 171 \AA~light curve were calculated in the same region, as marked with the white box in Figure 1 (A). By constructing a time-distance diagram along the arrow indicated by the dashed line in Figure 1 (A), one can clearly see continuous occurrence of weak plasma outflows after the fan-spine structure formation, as displayed in Figure 4 (B) and (C). Prior to the first ejection of jet, the weak plasma outflows exist during a period of approximately 77 minutes, with an average velocity of $\sim$134 $\speed$ and an ejected distance of $\sim$20 Mm. The homologous jets also propagate in the direction of this dashed line, and the average velocity of the three ejections is $\sim$230 $\speed$. Furthermore, the homologous jets propagate to a distance of up to $\sim$300 Mm, as shown in Figure 4 (B).  Additionally, intermittent ejections can also be observed between the jets, with ejections reaching approximately $\sim$100 Mm in distance at velocities of $\sim$244 $\speed$.
The velocities of the plasma outflows, homologous jets and intermittent ejections were calculated through the application of linear fitting to the height-time measurements, as displayed by the dashed lines in Figure 4 (B) and (C). 

\par
\subsection{Quasi-separatrix layer between two nulls}
Figure 5 (A) shows a bright plasma sheet near the null point of the north fan-spine structure. After superimposing the two null points on the AIA 131 \AA~images, we can see that the two null points are very close to the tips of the bright plasma sheet, respectively.  At 06:25 UT, the bright plasma sheet is obscured by the ejected material, as illustrated by the red dashed line in Figure 5 (B). During the phase of homologous jet eruptions, it becomes more prominent (see Figure 5 (C)$-$(D)). Such a sheet-like plasma structure most likely traces out a current sheet as suggested by previous observations \citep[e.g.,][]{2014ApJ...797L..14T,2016NatPh..12..847L,2016NatCo...711837X,2019ApJ...879...74C,2021ApJ...915...39H}.
From 06:52 UT to 06:58 UT, the bright plasma in the current sheet demonstrates an extension motion from the south null to the north null (also see the animation\_fig6), indicating a potential interaction between two fan-spine structures. To confirm this interaction, we further computed the 3D squashing factor Q using the code developed by \citet{2022ApJ...937...26Z}, with the calculation region being the same as the local potential field extrapolation. Figure 5 (E) shows the Q-factor distribution in the possible interaction region between the two fan-spine structures (the white box). Figure 5 (F) shows a 2D distribution of Q-factor on a vertical slice passing through both null points. It is clear that the two null points are connected by an elongated sheet-like quasi-separatrix layer (QSL), where the magnetic field connectivity remains continuous but changes with a sharp
gradient \citep[e.g.,][]{1996A&A...308..643D,1997A&A...325..305D,2002JGRA..107.1164T,2005LRSP....2....7L,2012A&A...541A..78P,2022LRSP...19....1P}. Such a QSL, with a high threshold of log(Q) $>$ 5, is a desirable place for the formation of strong current layer and the occurrence of 3D reconnection \citep[e.g.][]{2014SoPh..289..107D,2014ApJ...788...85V,2019ApJ...887..118C,2019ApJ...871..105Z,2023ApJ...953..148S}. In accordance with our EUV observations, this result suggests that a current sheet very likely forms along the sheet-like QSL that connects null points of the two fan-spine structures.

\par
\subsection{Generation of plasma jets}
Figure 6 shows the first and third jet ejections. Prior to the first jet ejection, a mini-filament appears beneath the newly formed fan-spine structure, as shown in Figure 6 (A). The overlying restraining field above the mini-filament is denoted by the white arrow in Figure 6 (A). At $\sim$06:34 UT, the overlying loop disappears, and the mini-filament itself is lifted upward. At $\sim$06:54 UT, the magnetic field structure of the mini-filament changes from a closed magnetic flux system to an open magnetic field. In the meantime, the twist contained in the mini-filament is released during the rotational motion of the jet. At $\sim$07:12 UT,  post-flare loops appear, as denoted by the white arrow in Figure 6 (D), implying that reconnection occurs beneath the mini-filament. Obviously, the production of this jet results from the eruptive mini-filament. The upward lifting and eruption of the mini-filament may be caused by the continuous magnetic flux emergence, and injection of magnetic helicity of shear term, as shown in Figure 2 (C) and Figure 4 (A). Alternatively, it could be due to the subsequent reconnection at this current sheet that weakens the restraining fields above the mini-filament. Figure 6 (E)$-$(G) displays the third jet accompanied by another pronounced mini-filament eruption. The white dashed line in Figure 6 (E) outlines the mini-filament. The change in the morphology of the mini-filament can be seen in Figure 6 (G). The newly formed field lines and post-flare loops after the reconnection are pointed by the blue and white arrows in Figure 6 (H). It should be noted that more than one filament are formed under this fan-spine magnetic field configuration. The three homologous jets originate from three different mini-filaments, two of which can be identified from Figure 6. These mini-filaments may be formed by the emerging arch filament systems (AFS) through photospheric rotation or converging motions \citep[e.g.,][]{2018SoPh..293...93C,2020ApJ...902....8C}. The AFS can carry the cold dense photospheric plasma and emerge to chromospheric or even coronal heights \citep[e.g.,][]{1967SoPh....2..451B,2018ApJ...853L..26H,2018ApJ...855...77S,2021ApJ...915...39H}.

\par
Figure 7, Figure 1 (B) and (F) display the different plasma motions from CHASE spectroscopic observations during the eruption of the homologous jets. During the initial phase of the first jet, the ejected plasma exhibits a blueshifted signature, while the circular flare ribbon and inner bright patch reveal redshifted signatures, as shown in Figure 1 (F) and Figure 7 (A).  After examining the timing of the red shifts in the circular flare ribbon and inner bright patch, we discovered that they became visible at $\sim$06:28 UT, approximately 13 minutes before the first peak of the AIA 171 \AA~light curve (see Figure 4 (A)). These features fade away around 3 minutes after the peak. As shown in Figure 7 (D), the H$\alpha$ line profiles sampled at the locations of the circular flare ribbon and inner bright patch all reveal an enhancement of the red wing absorption. The excess red wing absorptions at the circular ribbon and inner patch correspond to downward velocities of around 17$\speed$ and 13$\speed$, respectively. Subsequently, the entire fan of the fan-spine system exhibits redshifted signatures in the first decay phase of the AIA 171 \AA~light curve, as displayed in Figure 7 (B).  An excess absorption in the red wing of the H$\alpha$ line that corresponds to a speed of 19$\speed$ can be inferred from the green curve in Figure 7 (E). Additionally, during the third jet eruption, the remote brightening also displays red shifts, as denoted by the green arrows in Figure 7 (C) and Figure 1 (B). The redshifted feature at the remote brightening appears at 08:19 UT, three minutes after the third AIA 171 \AA~light curve peak at 08:16 UT, as displayed in Figure 4 (A). This feature persists for a duration of 10 minutes. The H$\alpha$ line profile shows a redshifted component with a speed of 15$\speed$. These results indicate the presence of redshifted signatures in the circular ribbon and inner patch (panel (A)) both before and during the first jet eruption. This phenomenon can be attributed to chromosphere condensation \citep[e.g.,][]{ 2015ApJ...811..139T,2017ApJ...848..118L,2023ApJ...954....7L}. While in Figure 7 (B), the redshifted signature is likely caused by the cooling plasma falling back to the solar surface along the dome shaped fan. Furthermore, the redshifted signature observed in the remote brightening is likely a result of chromospheric condensation and/or plasma downflows, as it persisted during and after the third jet eruption. Additionally, the Doppler velocities of the ejected plasma were estimated to be -18$\speed$, -5$\speed$, -12$\speed$ from Figure 7 (G), (H), and (I), respectively.

\par

\par

\section{Summary and Discussion} \label{sec:summ}
Using SDO/AIA and SDO/HMI observations, we have investigated the formation of fan-spine topology on the solar disk driven by magnetic flux emergence, and the subsequent jet production. We have identified two fan-spine structures and the two associated null points located at different heights. The detailed formation process of the north fan-spine structure and the accompanied plasma dynamics have been well observed and carefully analyzed. When the magnetic flux emerges, a set of small loops appears and undergoes interchange reconnection with the ambient open fields (plume), resulting in the generation of new field lines. As this process goes on, a structure that resembles a spider- or anemone-like fan dome starts to form. At the time when there is a sudden increase in magnetic flux, the magnetic null point is present at 4.1 Mm.  The newborn null point signifies the establishment of a typical fan-spine magnetic field configuration, whose topology exhibits a relatively axisymmetric morphology. After tracking the fan-spine null point with time, we find that the position of the magnetic null point displays a dynamic evolution, drifting eastward and then northward. The fan-spine structure progressively becomes non-axisymmetric as the null point drifts. Following the presence of null point, continuous plasma outflows appear and eject at an average velocity of 134$\speed$ for 77 minutes. This suggests that persistent gentle reconnection occurs at this magnetic null point. These plasma outflows are likely caused by the reconnection between the inner spine and the ambient open fields. Under the fan-spine magnetic field topology, mini-filaments form as a result of the persistent magnetic flux emergence and magnetic helicity injection. We also found that the nulls in the north and south fan-spine structures are connected by an elongated sheet-like QSL, in which a current sheet is most likely developed. Reconnection at the current sheet may eliminate the  overlying restraining fields above the mini-filaments. These processes result in the continuous eruptions of mini-filaments, which subsequently trigger three significant homologous jets. The average ejection velocity and distance for the three jets are 230$\speed$ and 300 Mm, respectively. With CHASE H$\alpha$ spectroscopic observations, we detected plasma motions during the homologous jet eruptions. The circular flare ribbon, inner bright patch, and remote brightening all exhibit redshifted signatures with velocities of a few tens$\speed$. The observed features in the red wing of the circular flare ribbon and inner patch are attributed to chromospheric condensation. While the redshifted signature in the remote brightening might be caused by chromospheric condensation and/or cooling downflows.

\par 
Recently, through the utilization of high-resolution EUV observations and the application of magnetic field extrapolation techniques, it has been revealed that various solar eruptions, including solar jets, occur frequently within the 3D fan-spine magnetic topology \citep[e.g.,][]{2012ApJ...760..101W,2019ApJ...885L..11S,2022ApJ...926L..39D}. These findings highlight the significance of the fan-spine magnetic configuration in understanding solar eruptive phenomena across different scales \citep[e.g.,][]{2017Natur.544..452W,2018ApJ...852...98W}. Most existing works primarily focused on solar eruptions within the fan-spine framework, however, there is very limited available literature on the formation of the fan-spine structure. \citet{2009ApJ...704..485T} suggested that a two-step reconnection process may play a role in the formation of a fan-spine system. Some previous works mentioned the possible links among flux emergence and fan-spine topologies \citep[e.g.,][]{2011ApJ...728..103L,2016ApJ...819L...3Z}, but none of them provided direct evidence to characterize their associations. In our study, we provide solid evidence demonstrating that the formation of a fan-spine structure is largely driven by magnetic flux emergence. The north null point emerges approximately 40 minutes after a significant increase in the magnetic flux emergence and the magnetic helicity injection. The weak plasma outflows are observed following the emergence of the null point, indicating continuous occurrence of the gentle reconnection at the null point.  Similarly, \citet{2023NatCo..14.2107C} found that magnetic reconnection occurred at a null point continuously at previously unresolved scales, providing evidence for reconnection at small-scale fan-spine structures. In our study, the duration of reconnection at the null point (77 minutes) is close to their total duration of 60 minutes, while the velocity of the plasma outflows ($\sim$134$\speed$) is larger than their velocity of $\sim$80$\speed$. Our observation also supports that the gentle reconnection is associated with the emerging flux that interacts with the nearby dominant opposite polarity. Furthermore, this continuous gentle reconnection appears to be caused by the inner spine of the fan-spine structure reconnecting with ambient open field lines at the null point.

\par
To date, many theoretical and MHD numerical studies have proposed two primary families of jet models to explain the generation of solar jets \citep{2016SSRv..201....1R,2021RSPSA.47700217S}. One is the emerging-reconnection jet scenario \citep[e.g.,][]{1977ApJ...216..123H,1992PASJ...44L.173S,1995Natur.375...42Y,2008ApJ...673L.211M,2013ApJ...769L..21A}, in which emerging magnetic flux reconnecting with ambient magnetic fields leads to jet ejections. This scenario can account well for the observed inverted Y-shaped structure of jets \citep[e.g.,][]{2013ApJ...771...20M,2018ApJ...854...92T}. Another model is the embedded-bipole model \citep[e.g.,][]{2009ApJ...691...61P,2010ApJ...714.1762P,2017Natur.544..452W,2018ApJ...852...98W}, in which an initial potential magnetic field with a bipole is embedded within a background field. The magnetic free energy is stored through photospheric rotation and shearing, and is then released through the magnetic reconnection process. As demonstrated by \citet{2015A&A...573A.130P}, the 3D null point fan-spine configuration is a favorable structure for energy storage and impulsive release. Our observation highlights the significance of magnetic flux emergence in the formation of the fan-spine topology and provides a general framework for explaining events where free energy is stored in the fan-spine system as a consequence of magnetic flux emergence. Initially, due to the interchange reconnection between the emerging flux and the ambient magnetic field, a spider- or anemone-like structure is formed. As simulated by \citet{2009ApJ...691...61P}, the important role of this configuration in their embedded-bipole model is to suppress the reconnection until a large amount of free energy has built up. Following the emergence of the fan-spine null point, a typical fan-spine topology is established, showing a relatively axisymmetric morphology. The research conducted by \citet{2015A&A...573A.130P,2016A&A...596A..36P} demonstrates that the simple forcing of a null point topology induces a low-intensity reconnection, such as the persistent plasma outflows observed in the initial phase of fan-spine structure formation in our observation. Furthermore, many flux emergence simulations interpreting reconnection events as coronal jets \citep[e.g.,][]{1995Natur.375...42Y,2008ApJ...673L.211M} may actually correspond to the plasma outflows resulting from the low-intensity reconnection (or gentle reconnection in our work), which are more closely related to the activation of the 3D null point fan-spine system than the genuine generation of coronal jets. 
\par
Subsequently, the null point drifts towards the emerging magnetic loops, and the drifting direction is consistent with that in the simulation conducted by \citet{2009ApJ...704..485T}. As a result of the drift of the null point, the fan-spine magnetic field configuration gradually becomes non-axisymmetric. \citet{2018ApJ...852...98W} discussed the intensity of the null point reconnection in the fan-spine framework, which depends on the inclination of the ambient magnetic field. When the magnetic field is highly inclined, it produces faster and more intense outflows. Our results also suggest that the obvious plasma ejection of the homologous jets occurs only in a non-axisymmetric fan-spine topology, which agrees with the simulation of \citet{2009ApJ...691...61P,2010ApJ...714.1762P} and \citet{2018ApJ...852...98W}. In the scenario of the non-axisymmetric topology, a thin current sheet is formed along the QSL (see Figure 5), in which microscale MHD instabilities may occur and result in a higher reconnection rate \citep[e.g.,][]{2006Natur.443..553D,2009Phys.Plasma....16..112102B,2010PhPl...17f2104H,2020RSPSA.47690867N}. This process likely leads to a faster and more efficient energy release, facilitating the production of homologous jets. The occurrence of the three homologous jets may be attributed to the energy accumulation reaching a certain threshold as evidenced by the formation of mini-filaments \citep[e.g.,][]{2018SoPh..293...93C,2020ApJ...902....8C}. These observational results align with the simulations carried out by \citet{2015A&A...573A.130P,2016A&A...596A..36P}, which demonstrate that the genuine generation of coronal jets (with significant untwisting motion) only occur when a sufficient amount of energy is built up. Similarly, the variation in reconnection intensity from slow to explosive has also been found during the occurrence of macrospicules \citep{2023ApJ...942L..22D}. Finally, there appears to be no significant plasma ejection after the end of the three homologous jet eruptions. Although there is still ongoing magnetic flux emergence in the ephemeral region, the background field has dissipated through reconnection. In the meantime, the magnetic null point has drifted to a non-emerging region. These lead to the end of eruptive phenomena.

\par
Although we have mainly focused on the formation of the north fan-spine structure and the accompanied plasma dynamics, our magnetic topology analysis reveals that a sheet-like QSL connects the nulls of the two fan-spine structures. Indeed, EUV imaging observations show that a current sheet is most likely formed along this sheet-like QSL, which spatially connects the south and north nulls. This suggests an interaction occurring between the two fan-spine structures. 
We have to mention that this interaction may play a role in triggering homologous jets. Because reconnection at the current sheet likely weakens the restraining field above the mini-filaments and facilitates eruptions of the mini-filaments and homologous jets. This process is similar to the classical breakout reconnection in the simulations of coronal jets from \citet{2017Natur.544..452W,2018ApJ...852...98W}.
Certainly, further observations and simulations are required to confirm the interactions among multiple fan-spine structures. In particular, considering the 3D reconnection processes in the corona, the contribution of these multiple fan-spine interactions to coronal heating needs to be investigated more extensively.


\acknowledgments
The authors are grateful for the anonymous referee's valuable comments and suggestions. This work is supported by National Key R\&D Program of China No 2021YFA0718600, the Natural Science Foundation of Beijing (1244053), and National Postdoctoral Programs (GZC20230097, 2023M740112). H. C. Chen was supported by the NSFC grant 12103005 and Yunnan Key Laboratory of Solar Physics and Space Science under the number YNSPCC202210. Z. Y. Hou was supported by NSFC grant 12303057. Y. D. Shen was supported by the NSFC grant (12173083), the Yunnan Science Foundation for Distinguished Young Scholars (202101AV070004). C.Li was supported by the NSFC grant (12333009). The authors are grateful for the data provided by the SDO, CHASE science teams. The CHASE mission is supported by China National Space Administration.

\begin{figure}[t]    
\centerline{\includegraphics[width=1\textwidth,clip=]{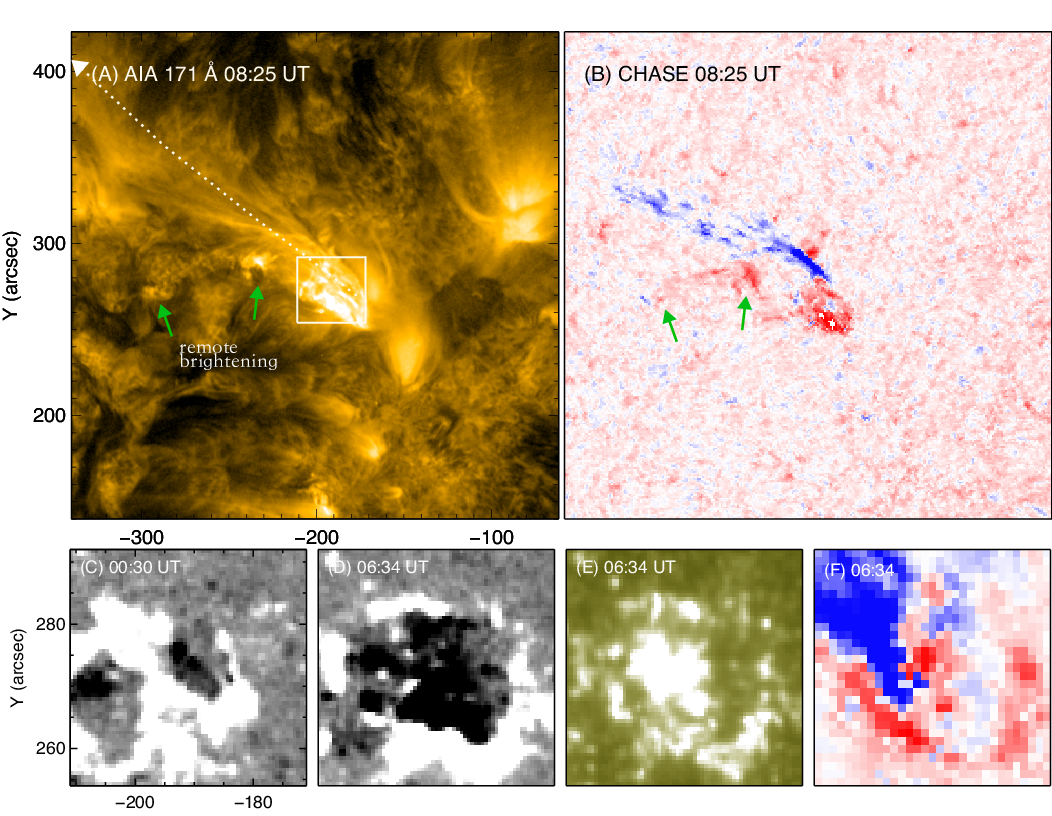}}
\caption{Overview of the ephemeral region. (A)$-$(B): AIA 171 \AA~and CHASE Doppler images showing the third jet. The green arrows denote the remote brightenings associated with the jet. The FOV of (C)$-$(F) is indicated by the white box in (A). (C)$-$(D): a time sequence of HMI LOS magnetograms showing the variations of the negative and positive magnetic fields, scaled at $\pm$50 G, respectively. (E)$-$(F): AIA 1600 \AA~and Doppler images showing the circular ribbon during the jet. 
}
\label{fig1}
\end{figure}
 
\begin{figure}[b]    
\centerline{\includegraphics[width=1\textwidth, clip=]{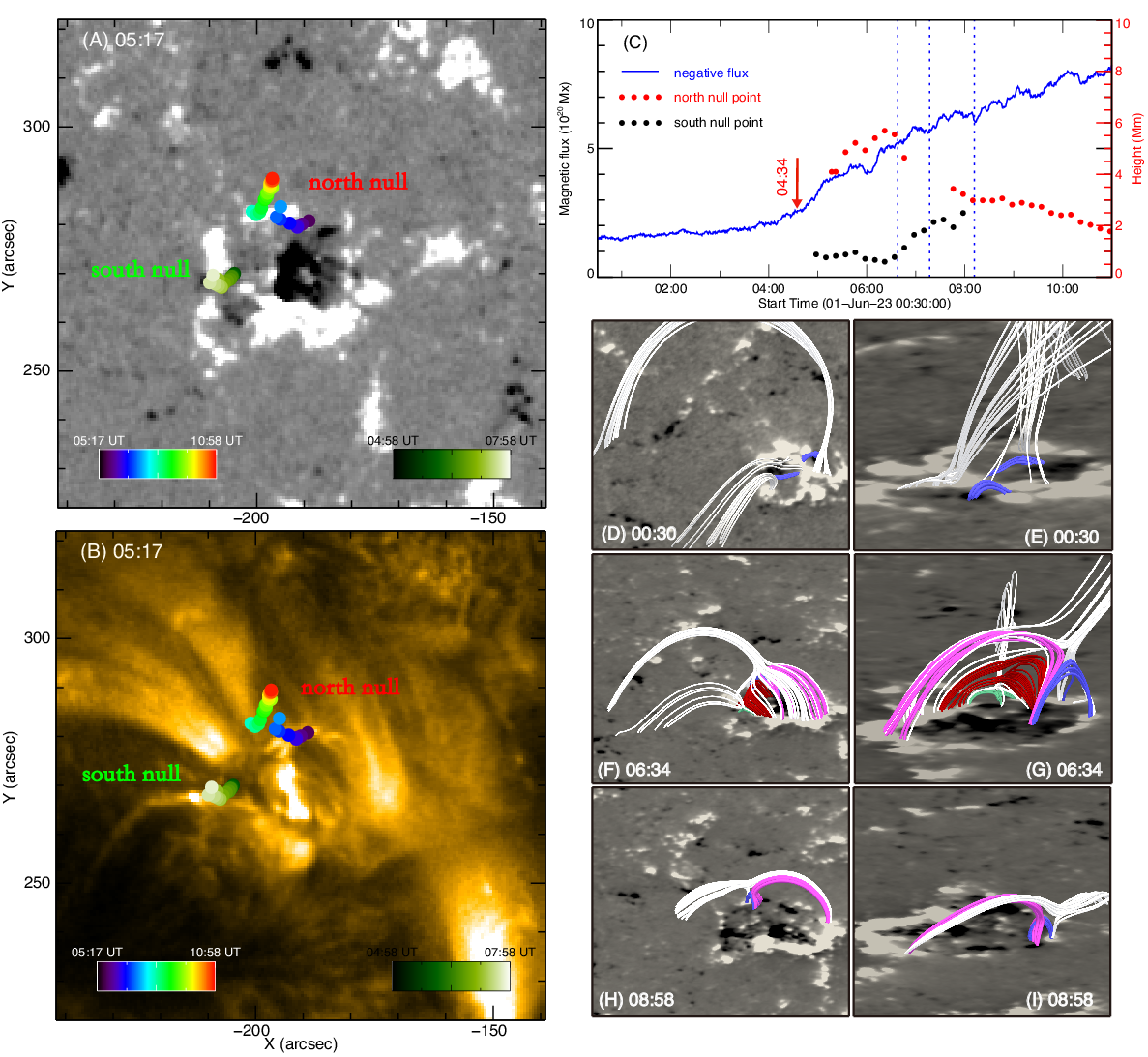}}
\caption{(A)$-$(B): evolution of the north and south null point positions projected on an HMI magnetogram and an AIA 171 \AA~image, respectively.
(C): the evolution of the negative magnetic flux (blue curve). The heights of the north null point (red solid circles) and south null point (black solid circles) . The negative magnetic flux is calculated in the FOV marked by the white box in Figure 1 (A). The three vertical dashed lines indicate the occurrences of the three homologous jets. (D), (F), and (H): the changes of the two fan-spine structures overplotted on sequential magnetograms. (E), (G), and (I): close-up views of the morphological evolution of the system from another perspective. The blue lines represent the emerging loops in (D)$-$(E). In (F)$-$(G), the white lines indicate the outer spines, the blue and pink lines represent the inner spine lines of north fan-spine structure, and the green lines depict the inner spine lines of south fan-spine structure. \textbf{The red lines represent field lines traced from the vicinity of the south null.} An animation is available, the duration of the animation is 4 s.
}
\label{fig2}
\end{figure}

\par
\begin{figure}    
\centerline{\includegraphics[width=1\textwidth,clip=]{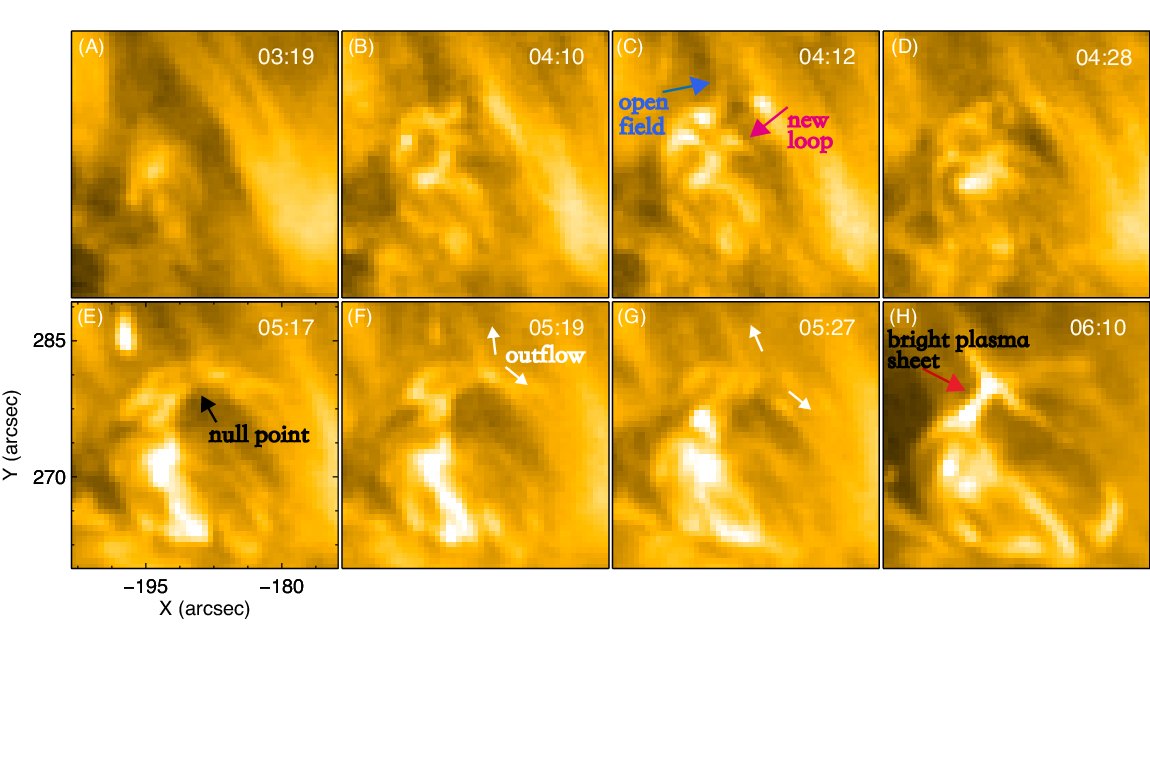}}
\caption{AIA 171 \AA~images displaying the formation process of the fan-spine system. The blue and pink arrows denote the newly formed open field lines and loops in (C). The black arrow denotes the location of the null point in (E). The white arrows point to the plasma outflows moving along magnetic field lines in (F)$-$(G). The red arrow points to a bright plasma sheet in (H). An animation of this formation process is available, showing the interaction between the emerging magnetic loops with surrounding opposite-polarity open field lines from 03:15 UT to 06:16 UT. The duration of the animation is 8 s.
}

\label{fig3}
\end{figure}

\begin{figure}    
\centerline{\includegraphics[width=0.75\textwidth,clip=]{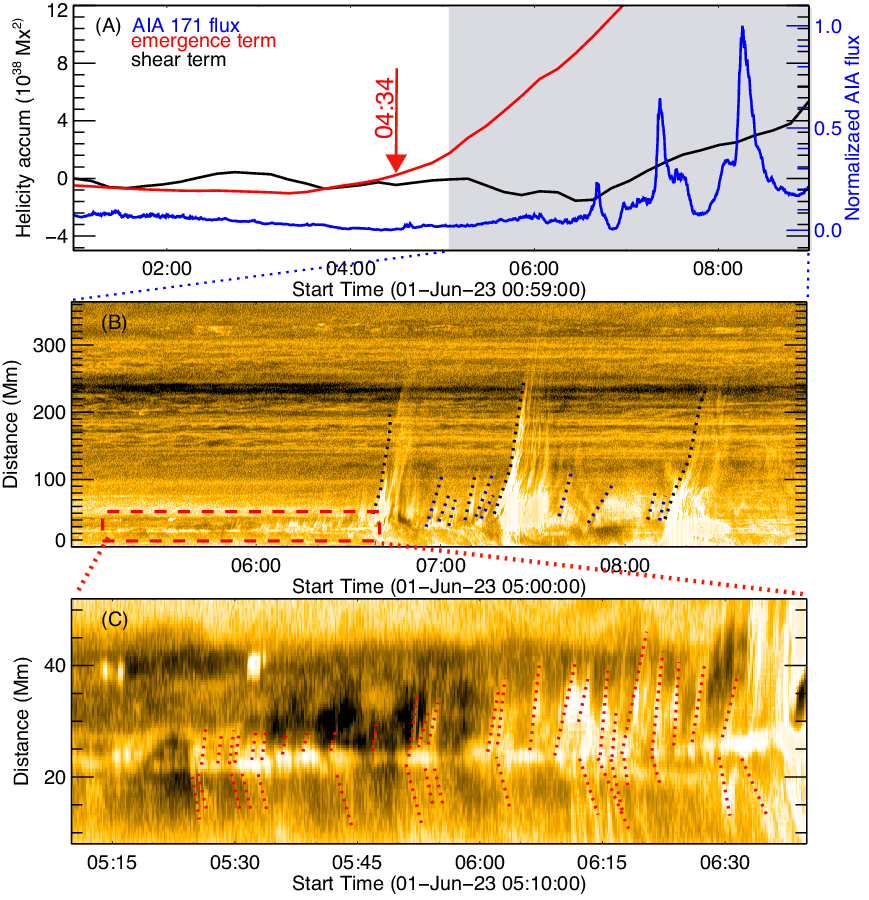}}
\caption{The accumulation of helicity injection and the AIA 171 \AA~light curve calculated within the white box shown in Figure 1 (A) from 00:59 UT to 09:00 UT. In (A), the black and red curves represent the shear and emergence terms of the magnetic helicity, respectively. The temporal evolution of AIA 171 \AA~flux is indicated by the blue curve. The red arrow points to the sharp increase in the injected helicity, respectively. (B) time–distance plot along the direction of the white dashed line in Figure 1 (A). (C) close-up view of the red dashed box in (B).
}
\label{fig4}
\end{figure}

\begin{figure}    
\centerline{\includegraphics[width=0.8\textwidth,clip=]{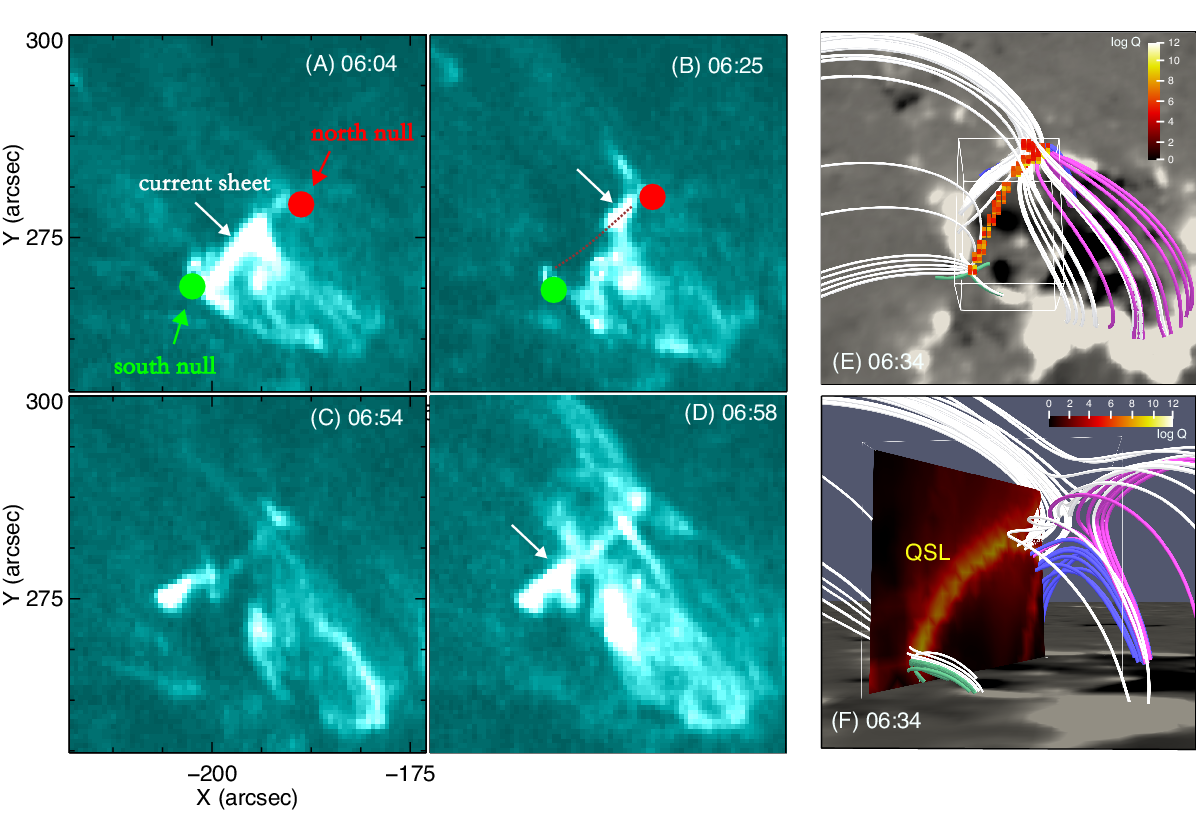}}
\caption{Sequential images of AIA 131 \AA~displaying the current sheet. (A)$-$(B): the solid red and green dots represent the positions of the north and south magnetic null points, respectively. (E)$-$(F): extrapolated magnetic field lines over the magnetogram as viewed from two different perspectives. In (E) locations with log (Q) $>$ 5 within the white box are illustrated. In (F), a 2D distribution of factor Q on a vertical slice passing through both nulls is presented.
}
\label{fig5}
\end{figure}

\begin{figure}    
\centerline{\includegraphics[width=0.8\textwidth,clip=]{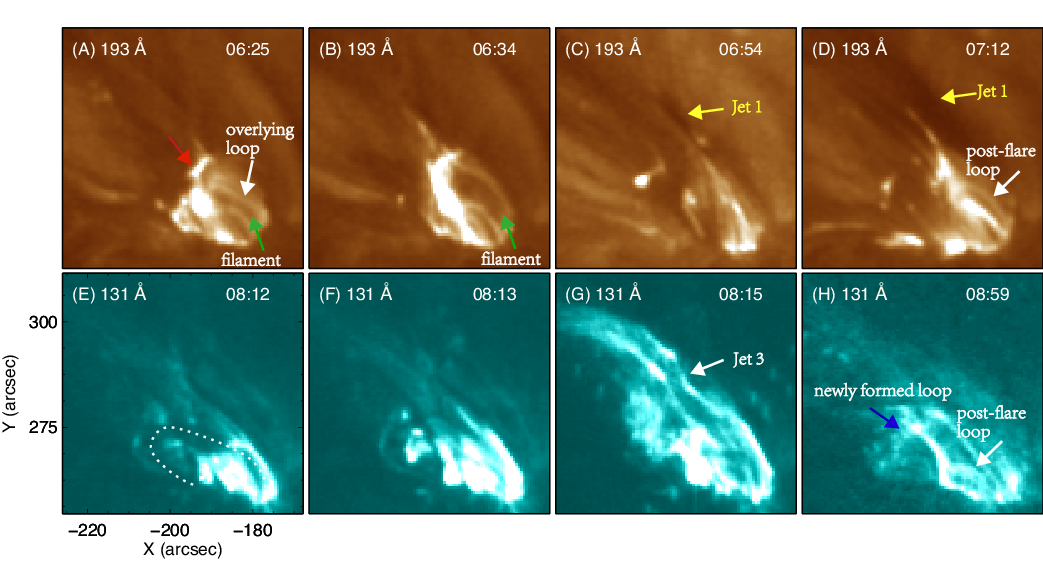}}
\caption{Generation of the homologous jets seen from the AIA 193 and 131 \AA~images. (A)$-$(D) and (E)$-$(H) show the eruption processes of the first and third jets, respectively. In (A), the green, red, and white arrows point to the filament, current sheet, and overlying loops above the filament, respectively. The yellow and white arrows in (C) and (G) denote the first and third jets, respectively. In (D) and (H), the white arrow denotes the post-flare loop, and the blue arrow points to the newly formed loop. An animation of the production process of the homologous jets is available. The duration of the animation is 13 s, ranging from 05:59 UT to 08:59 UT.
}
\label{fig6}
\end{figure}

\begin{figure}    
\centerline{\includegraphics[width=0.8\textwidth,clip=]{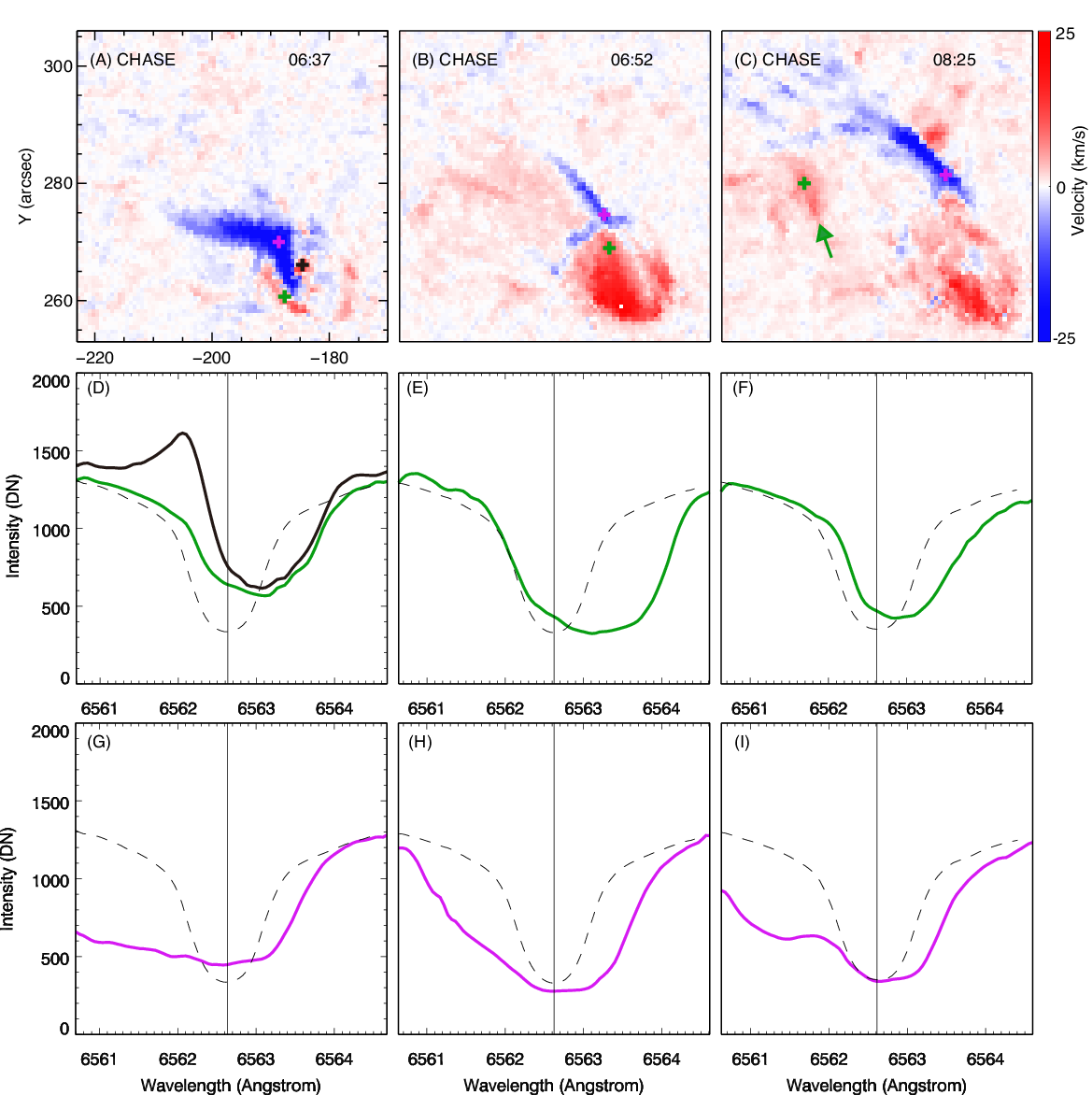}}
\caption{The Doppler maps and spectra obtained from CHASE showing the plasma motions of the homologous jets at different times. (D)$-$(I) represent the line profiles of selected points in different regions, marked with cross symbols in (A)$-$(C). The line profiles in green and pink show excess absorptions in the red and blue wings, respectively. The dashed line represents the average H$\alpha$ profile. The solid black line indicates the line profile of the inner bright patch. 
}
\label{fig7}
\end{figure}

\end{document}